# Deep Learning based Emotion Recognition System Using Speech Features and Transcriptions

Suraj Tripathi[1], Abhay Kumar[1*], Abhiram Ramesh[1*], Chirag Singh[1*], Promod Yenigalla[1]

[1]Samsung R&D Institute India – Bangalore
suraj.tri@samsung.com, abhay1.kumar@samsung.com,
a.ramesh@samsung.com c.singh@samsung.com, promod.y@samsung.com

**Abstract.** This paper proposes a speech emotion recognition method based on speech features and speech transcriptions (text). Speech features such as Spectrogram and Mel-frequency Cepstral Coefficients (MFCC) help retain emotion-related low-level characteristics in speech whereas text helps capture semantic meaning, both of which help in different aspects of emotion detection. We experimented with several Deep Neural Network (DNN) architectures, which take in different combinations of speech features and text as inputs. The proposed network architectures achieve higher accuracies when compared to state-of-the-art methods on a benchmark dataset. The combined MFCC-Text Convolutional Neural Network (CNN) model proved to be the most accurate in recognizing emotions in IEMOCAP data. We achieved an almost 7% increase in overall accuracy as well as an improvement of 5.6% in average class accuracy when compared to existing state-of-the-art methods.

## 1   INTRODUCTION

A majority of natural language processing solutions, such as voice-activated systems, chatbots, etc. require speech as input. Standard procedure is to first convert this speech input to text using Automatic Speech Recognition (ASR) systems and then run classification or other learning operations on the ASR text output. Kim *et al.* [1] trained CNNs on top of pre-trained word vectors for sentence-level classification and achieved state-of-the-art results on multiple benchmarks. Zhang *et al.* [2] used character level CNNs for text classification and showed comparable results against traditional models such as bag-of-words, n-grams and their TF-IDF variants, word-based ConvNets and Recurrent Neural Networks (RNN).

Human-computer interaction can get more interactive and personalized as computers improve in predicting the current emotional state of the human speaker, helping them in distinguishing different contextual meanings of the same word. ASR resolves varia-

---

\* equal contribution

42tions in a speech from different users using probabilistic acoustic and language models [3], which results in speech transcriptions being speaker independent. This might be good enough for most applications, but is an undesired result for systems which rely on knowing the intended emotion to function correctly. State-of-the-art ASR systems produce outputs with high accuracy but end up losing a significant amount of information that suggests emotion from speech. This gap has resulted in Speech-based Emotion Recognition (SER) systems becoming an area of research interest in the last few years.

Speech is one of the natural ways for humans to express their emotions. Moreover speech is easier to obtain and process in real time scenarios, which is why most applications that depend on emotion recognition work with speech. A typical SER system works on extracting features such as spectral features, pitch frequency features, formant features and energy related features from speech, following it with a classification task to predict various classes of emotion [4] [5]. Bayesian Network model [6] [7], Hidden Markov Model (HMM) [8], Support Vector Machines (SVM) [9], Gaussian Mixture Model (GMM) [10] and Multi-Classifier Fusion [11] are few of the techniques used in traditional classification tasks.

Since the last decade, Deep Learning techniques have contributed significant breakthroughs in natural language understanding (NLU). Deep Belief Networks (DBN) for SER, proposed by Kim *et al.* [12] and Zheng *et al.* [13], showed a significant improvement over baseline models [5] [11] that do not employ deep learning, which suggests that high-order non-linear relationships are better equipped for emotion recognition. Han *et al.* [14] proposed a DNN-Extreme Learning Machine (ELM), which uses utterance-level features from segment-level probability distributions along with a single hidden layer neural net to identify utterance level emotions, although improvement in accuracies were limited. Fayek *et al.* [15] made use of deep hierarchical architectures, data augmentation and regularization with a DNN for SER, whereas Zheng *et al.* [16] used Spectrograms with Deep CNNs. Vladimir *et al.* [17] trained DNNs on a sequence of acoustic features calculated over small speech intervals along with a probabilistic-natured CTC loss function, which allowed the consideration of long utterances containing both emotional and unemotional parts and improved recognition accuracies. Lee *et al.* [4] used a bi-directional LSTM model to train the feature sequences and achieved an emotion recognition accuracy of 62.8% on the IEMOCAP [18] dataset, which is a significant improvement over DNN-ELM [14]. Satt *et al.* [19] used deep CNNs in combination with LSTMs to achieve better results on the IEMOCAP dataset.

In recent years researchers have been looking into the use of multimodal features for emotion recognition. Tzirakis *et al.* [20] proposed an SER system that uses auditory and visual modalities to capture emotional content from various styles of speaking. Zadeh *et al.* [21] proposed a Tensor Fusion Network, which learns intra-modality and inter-modality dynamics end-to-end, ideal for the volatile nature of language online. Ranganathan *et al.* [22] experimented with Convolutional Deep Belief Networks (CDBN), which learn salient multimodal features of expressions, to achieve good accuracies.

In this work, we propose a robust technique of emotion classification using speech features and transcriptions. The objective is to capture emotional characteristics using speech features, along with semantic information from text, and use a deep learning based emotion classifier to improve emotion detection accuracies. We present different deep network architectures to classify emotion using speech features and text. The main contributions of the current work are:
- Proposed a CNN model for emotion classification using Speech features (MFCC, Spectrogram)
- Proposed a CNN model for emotion classification using both speech features (MFCC, Spectrogram) and transcriptions

## 2  PROPOSED METHODS

In this paper, we consider speech transcriptions along with its corresponding speech features – Spectrogram and MFCC, which together provide a deep neural network both semantic relationships and the necessary low-level features required to distinguish among different emotions accurately. Experiments have been performed on speech transcriptions and speech features independently as well as together in an attempt to achieve accuracies greater than existing state-of-the-art methods. Different combinations of inputs have been used in different DNN architectures, the details of which have been discussed in the following section.

### 2.1  CNN Model based on Text (Model 1)

Speech transcriptions are widely used in various sentiment analysis applications [1] [2]. In emotion detection, it is important for a model to understand the context of the utterance correctly to be able to predict its intent accurately. Let us take the word "good" for example. Used on its own it is hard to know the context in which this word has been used. Although the word "good" implies something positive, it could have possibly been part of a larger conversation and could have been used in a sarcastic way. The following is an example utterance from the IEMOCAP data, "It'll be good. Wow! That's great." Here we can clearly see, without any ambiguity, the context in which the word "good" is used. DNNs are good at learning such contextual information and aid in considering the bigger picture rather than focusing on individual terms. The CNN model described in this section takes speech transcriptions, in the form of word embeddings, as input to detect emotion. CNNs can directly be applied to word embeddings without prior information on their semantic contexts. This strongly suggests that language can be interpreted as a signal no different from other signals.

Word embedding describes feature learning and language modelling techniques where words and phrases from a given vocabulary are mapped to vectors or real numbers. As word embeddings are trained to predict the surrounding words in the sentence, these vectors can be viewed as representing the contexts in which words are distributed and can thus be considered an ideal representation for DNNs. Fig. 1 shows the word embedding plot for three classes of emotions (Happiness, Sadness, and Anger) in two dimensions. The plot clearly shows words either synonymous with each other or used in similar contexts, such as *felicity*, *cheerful* and *like* are grouped together. Word embeddings close to each other could have the same emotion associated with it and a DNN can pick up on these contextual patterns during training.

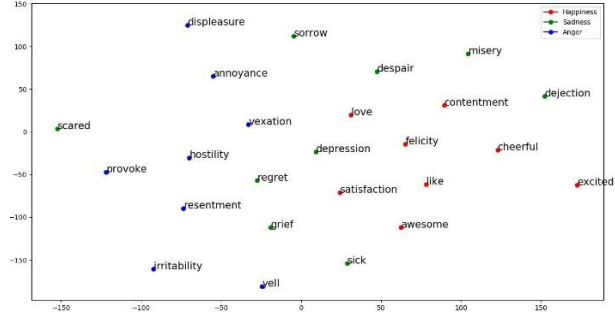

**Fig. 1.** t-SNE visualization of word embedding vectors

Fig. 2 represents the architecture of the text-based CNN model used in our experiments. Transcription sequences (embedded vectors), which is the input to this model, are convolved with kernels of different sizes. The maximum number of words in any given utterance is set to 128, which covers almost the entirety of the IEMOCAP dataset. One feature from each of the different convolutional layers is picked by the max-pool layer. These features are fed to a single FC layer. Finally, a softmax layer is used to perform classification. We experimented with batch-normalization, a technique which helps prevent the model from over-fitting and also from being too sensitive to the initial weight, and also varied dropout rates from 0.25 to 0.75. An improvement in convergence rate is also observed with the use of batch-normalization.

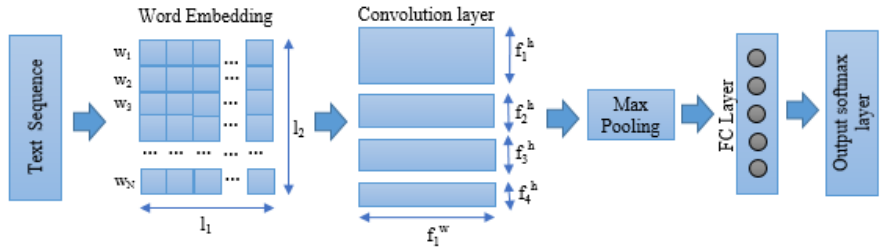

**Fig. 2.** Text-based CNN model

Word embeddings, from Google Word2Vec [23], of size 300 are used in this experiment. Filter widths are kept consistent with the length of the word embedding, but the heights vary between 12, 8, 6 and 3. This is done in an attempt to take advantage of filters of different sizes. These filters are applied in parallel, and the coefficients are tuned as a part of the training process.

## 2.2 CNN Model based on Speech Features

Recent research into speech processing has shown the successful application of deep learning methods and concepts such as CNNs and Long Short Term Memory (LSTM) cells on speech features. CNNs have been shown, by extensive research, to be very useful in extracting information from raw signals in various applications such as speech recognition, image recognition, etc. In our work, we use Spectrograms and MFCCs, as they are commonly used to represent speech features, along with CNNs for emotion detection.

**CNN Model with Spectrogram input (Model 2A)**

A spectrogram is a representation of speech over time and frequency. 2D convolution filters help capture 2D feature maps in any given input. Such rich features cannot be extracted and applied when speech is converted to text and or phonemes. Spectrograms, which contain extra information not available in just text, gives us further capabilities in our attempts to improve emotion recognition.

The following model uses Mel-frequency Spectrogram as input to a 2D CNN. Spectrograms are generated when Short Term Fourier Transform (STFT) is applied on windowed audio or speech signal. The audio is sampled at 22050Hz. Windowing is then carried out on each audio frame using a "*hann*" window of length 2048. Fast Fourier Transform (FFT) windows of length 2048 are then applied on the said windowed audio samples with an STFT hop-length equal to 512. The obtained Spectrogram magnitudes are then mapped to the Mel-scale to get Mel-spectrograms. 128 Spectrogram coefficients per window are used in this model. The Mel-frequency scale puts emphasis on the lower end of the frequency spectrum over the higher ones, thus imitating the perceptual hearing capabilities of humans. We used the `librosa` python package, along with the above mentioned parameters, to compute the Mel-spectrograms. A sample Spectrogram corresponding to audio "*I cannot... you are not here by choice. Nobody would ride this bus by choice*." is shown below in Fig. 3.

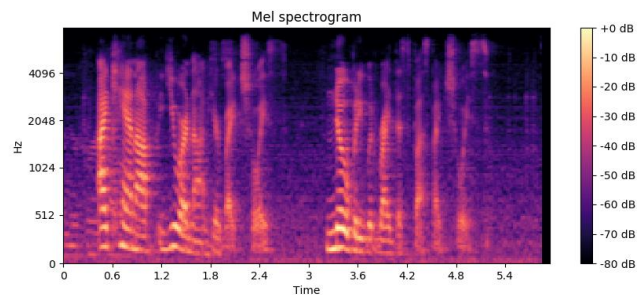

**Fig. 3.** Spectrogram of a sample audio file

In our CNN model, we take Spectrogram input with a maximum image width of 256 (number of windows). Since our dataset has audios of varying lengths, we trim long duration audio files to a fixed duration (6 seconds), which covers 75 percentile of all audio data samples of the dataset. This decision was made under the assumption that the frequency variations that characterize the emotionality of the speech data will be present throughout the dialogue and hence will not be lost by this reduction in length.

Fig. 4 details the 2D CNN architecture used to detect emotion using Spectrograms. A set of 4 parallel 2D convolutions are applied on the Spectrogram to extract its features. The input shape of the Spectrogram image is 128 x 256 (number of Mels x number of windows). 200 2D-kernels are used for each of the parallel convolution steps. Figuring out the optimal kernel size is a difficult and time taking task, which may depend on several factors all which cannot be clearly defined. To prevent choosing one single kernel size that could possibly be sub-optimal we decided to use kernels of different sizes, each of which is fixed for a single parallel path, to take advantage of the different patterns picked up by each kernel. The sizes of each of the kernels in their respective parallel CNN paths are 12 x 16, 18 x 24, 24 x 32, and 30 x 40. The features generated in the said convolution layers are then fed to their respective max-pool layers, which extracts 4 features from each filter as the pool size is exactly half along the width and height of the convolution output. The extracted features are fed to the Fully Connected (FC) layer. This model makes use of two FC layers of sizes 400 and 200. Batch normalization is applied to both the FC layers. We experimented with dropout rates varying between 25% and 75% for the first FC layer but excluded it completely from the second FC layer. The activation function used in the convolutional layers and the first FC layer is the Rectified Linear Unit (ReLU). The output of the last FC layer is then fed to a Softmax layer, which classifies the input speech signal among 4 different emotion classes. "Adadelta" is the optimization technique used during training.

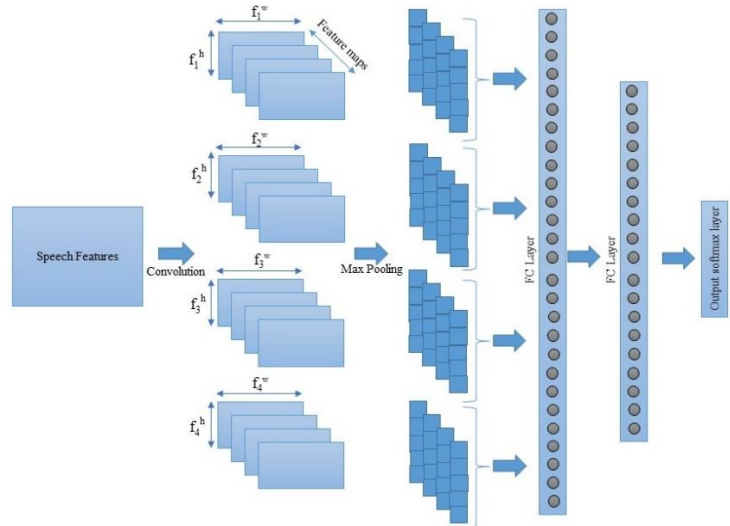

**Fig. 4.** Spectrogram/MFCC based CNN model

We experimented with another variant of this model (refer Model 2B). The architecture of this model is quite similar to Model 2A but is augmented with four additional parallel convolution layers. Spectrograms, which have been down-sampled by 2 in both time and frequency, are taken as the input. The convolutional filters in this model are kept as is in Model 2A. This experiment was conducted in an attempt to extract even higher level features as compared to the ones in Model 2A. The performance of this model is discussed in Table 1.

**CNN Model with MFCC input (Model 3)**

Mel Frequency Cepstrum (MFC), refer Fig. 5, is a representation of the Short-Term Power Spectrum of sound. It is based on a linear cosine transform of a log power spectrum on a non-linear Mel-scale of frequency. As MFCC is a popular speech feature widely used in various speech processing applications, we decided to use the same in our experiments on emotion detection.

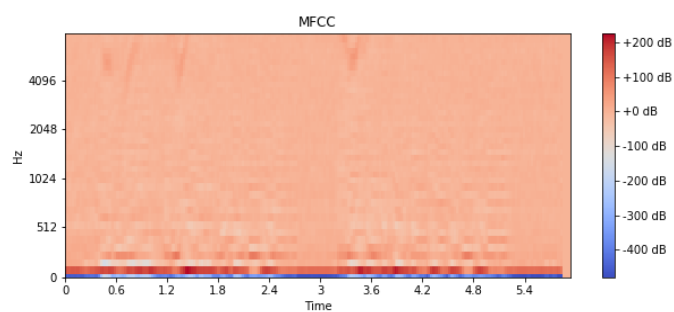

**Fig. 5.** MFCC of the audio file being referred to in Fig. 3

The hyper-parameters and the python package (`librosa`) used for MFCC generation are similar to the ones described for Spectrogram generation. The only difference is that 40 MFCCs per window are generated compared to the earlier mentioned 128 Spectrogram coefficients per window.

This model also consists of 4 sets of parallel convolutional layers, followed by max-pooling layers and 2 more FC layers, similar to the one described in the previous section. As the input size is different to that of Model 2A and 2B we experimented with kernels of different sizes and eventually chose the set of 4 x 6, 6 x 8, 8 x 10 and 10 x 12 kernels for this model.

### 2.3 Combined CNN Model based on both Text and Speech Features

We experimented with 3 different combined models in an attempt to bring together different strengths offered by each of Spectrogram, MFCC and speech transcriptions. Since inputs are different and thus of different dimensions, we use separate CNN channels, as represented in Fig. 6, for each. The models are:

— Spectrogram and MFCC model (Model 4A)
— Spectrogram and text model (Model 4B)
— MFCC and text model (Model 4C)

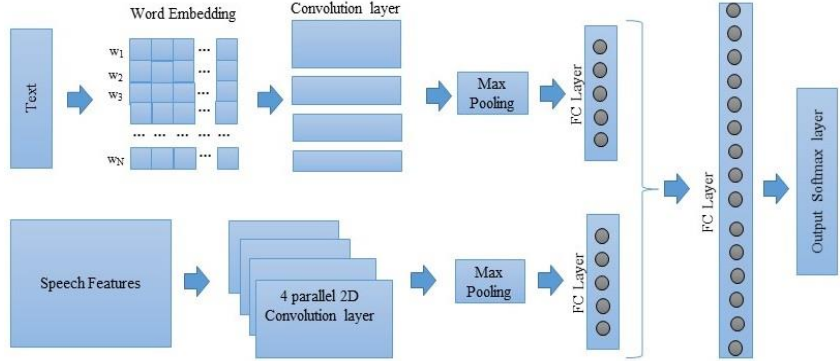

**Fig. 6.** Representative CNN architecture

In Model 4A the Spectrogram channel consists of 4 parallel 2D-CNN layers with kernels of different sizes. Like the Spectrogram channel the MFCC channel also consists of 4 parallel 2D-CNN layers. Outputs from both the channels are fed to one FC layer each. The outputs of both the FC layers, after normalization, are concatenated and fed to the 2nd FC layer. The final step is to feed the outputs of the last FC layer to a softmax layer. Architectures of Model 4B and 4C are similar to that of Model 4A. Both the models have a text channel that takes word embeddings as input but alternate between Spectrogram and MFCC as the speech feature.

## 3   DATASET

We used the University of Southern California's Interactive Emotional Motion Capture (USC-IEMOCAP) database in this work. The IEMOCAP corpus comprises of five sessions where each session includes the conversation between two people, in both scripted and improvised topics and their corresponding labeled speech text (both phoneme and word level). Each session is acted upon and voiced by both male and female voices to remove any gender bias. The audio-visual data thus collected is then divided into small utterances of length varying between 3-15 seconds, which are then labelled by evaluators. Each utterance is evaluated by 3-4 assessors. The assessors had the option of labelling every utterance among 10 different emotion classes (neutral, happiness, sadness, anger, surprise, fear, disgust frustration, excited, other). In our experiments, we have considered only 4 of them (anger, excitement (happiness), neutral and sadness) so as to remain consistent with earlier research. We chose utterances where at least 2 experts were in agreement with their decision and only used improvised data, again being consistent with prior research, as the scripted text shows a strong correlation with labeled emotions and can lead to lingual content learning, which can be an undesired side effect. The final experimental dataset extracted from the original IEMOCAP data comprised of 4 classes named Neutral (48.8% of the total dataset),

Happiness (12.3%), Sadness (26.9%) and Anger (12%). As there is data imbalance between different emotional classes, we present our results on overall accuracy, average class accuracy and also show the confusion matrix (refer to Table 2).

## 4    EVALUATION AND DISCUSSION

### 4.1    Emotion classification on standard dataset

In congruence with previous research efforts, we show the effectiveness of the proposed methods for emotion detection with our benchmark results on IEMOCAP dataset. We have used stratified K-fold for splitting training and test data. Some recent results on emotion classification are presented in Table 1, along with our 5-fold cross-validation experimental results. Both overall and class accuracies are presented for better comparison, where overall accuracy is measured based on total counts irrespective of classes, and class accuracy is the mean of accuracies achieved in each class.

**Table 1.** Comparison of accuracies

| Methods | Input | Overall Accuracy | Class Accuracy |
|---|---|---|---|
| Lee [4] | Spectrogram | 62.8 | 63.9 |
| Satt [19] | Spectrogram | 68.8 | 59.4 |
| Model 1 | Text | 64.4 | 47.9 |
| Model 2A | Spectrogram | **71.2** | 61.9 |
| Model 2B | Spectrogram | **71.3** | 61.6 |
| Model 3 | MFCC | **71.6** | 59.9 |
| Model 4A | Spectrogram & MFCC | **73.6** | 62.9 |
| Model 4B | Text & Spectrogram | **75.1** | **69.5** |
| Model 4C | Text & MFCC | **76.1** | **69.5** |

A text-only based CNN model fails to capture all the low-level features of speech signals and thus does not achieve a very high emotion detection accuracy. As mentioned above the combined MFCC and Spectrogram based CNN model achieves an overall emotion detection accuracy improvement close to 4% over existing state-of-the-art methods. The combined Text-MFCC model performs even better and beats the benchmark class accuracy by 5.5% and the overall accuracy by close to 7%. As mentioned earlier, the IEMOCAP data is not well balanced with respect to the amount of data amongst various classes of emotion. In Table 2 we present the confusion matrix, which shows misclassifications between each pair of emotion classes in Model 4C. From this table, we can observe that Neutral and Sadness classes have a high detection rate whereas Happiness and Anger classes suffer more misclassifications. This bias in classification accuracy could possibly be attributed to language experts more easily being able to identify sadness and neutral emotions in speech, whereas the task of labelling the other emotions could have been more subjective.

The published state-of-the-art emotion detection accuracy on data from the IEMOCAP corpus, to the best of our knowledge, is given in [19] and is based on the same evaluation setup used by us; it reports 68.8% and 59.4% for overall and class accuracies respectively. Jin *et al.* [5] and Tripathi *et al.* [24] report overall accuracies of 69.2% and 71.04% respectively, but the parameters on which data was selected is different to that of ours.

Table 2. Confusion Matrix in Percentage on the Model 4C

| Class Labels | Prediction | | | |
|---|---|---|---|---|
| | Neutral | Happiness | Sadness | Anger |
| Neutral | **81.30** | 4.74 | 11.68 | 2.45 |
| Happiness | 37.44 | **49.24** | 10.76 | 2.54 |
| Sadness | 13.94 | 1.08 | **84.06** | 0.89 |
| Anger | 30.43 | 3.84 | 2.29 | **63.41** |

### 4.2 Discussion

The text-only model provides important semantic relationships necessary to detect the intended emotion in an utterance. But it doesn't beat the state-of-the-art results, possibly due to the fact that converting speech to text results in the loss of very important low-level features which would have otherwise been vital for emotion detection. Speech features based models seem to perform better in our experiments as they contain information over both time and frequency, which when convolved with 2D kernels captures 2D feature maps much richer in information. The Spectrogram model referred to in Model 2A gives an overall accuracy of 71.2%, whereas its further complex variant referred to in Model 2B gives an emotion detection accuracy of 71.3%. The MFCC based model also gives a comparable result to the previously described models, coming in with an accuracy of 71.6%, an almost 2.5% improvement over existing state-of-the-art results. A sharp rise in accuracy can be seen when the CNN models get multiple inputs, the reason possibly being that different types of inputs offer specific features, all of which are needed to improve emotion detection. Model 4A, which takes in both Spectrogram and MFCC, gives an accuracy of 73.6%. Model 4B and 4C, which work with Text-Spectrogram and Text-MFCC respectively, provide an overall accuracy of 75.1% and 76.1%, the latter being a 7% improvement over benchmark results.

## 5 CONCLUSIONS

In this paper, we have proposed multiple CNN based architectures to work with speech features and transcriptions. Speech features based 2D CNN model provides better accuracy relative to state-of-the-art results, which further improves when combined with text. The combined Spectrogram-MFCC model results in an overall emotion detection accuracy of 73.1%, an almost 4% improvement to the existing state-of-the-art methods. Better results are observed when speech features are used along with speech transcriptions. The combined Spectrogram-Text model gives a class accuracy of 69.5% and an overall accuracy of 75.1% whereas the combined MFCC-Text model also gives a class accuracy of 69.5% but an overall accuracy of 76.1%, a 5.6% and an almost 7% improvement over current benchmarks respectively. The proposed models can be used for emotion-related applications such as conversational chatbots, social robots, etc. where identifying emotion and sentiment hidden in speech may play a role in the better conversation.